\documentclass[a4paper,fleqn,usenatbib]{mnras}
\usepackage{newtxtext,newtxmath}
\usepackage[T1]{fontenc}
\usepackage{ae,aecompl}
\usepackage{epsfig}
\usepackage{amsmath, amsfonts, epsfig, xspace}
\usepackage{natbib}
\usepackage{deluxetable}
\usepackage{rotating}
\usepackage{graphicx}
\usepackage{caption}
\usepackage{longtable}
\usepackage{multicol}
\usepackage{booktabs}
\usepackage{dcolumn}
\usepackage{longtable}
\newcolumntype{d}[1]{D{.}{.}{#1}}
\usepackage[dvisvgm, usenames, dvipsnames]{xcolor}
\usepackage[labelfont=bf]{caption}
\usepackage{subfig}
\usepackage{floatrow}
\usepackage{float}
\usepackage[morefloats=200]{morefloats}
\usepackage{threeparttablex}
\def\aj{{AJ}}%
\def\araa{{ARA\&A}}%
\def\apj{{ApJ}}%
\def\apjl{{ApJ}}%
\def\apjs{{ApJS}}%
\def\aap{{A\&A}}%
\def\aaps{{A\&AS}}%
\def\mnras{{MNRAS}}%
\def\pasp{{PASP}}%
\title[Persistence of blazar state in FSRQs]{Persistence of blazar state in flat-spectrum radio quasars}
\author[Krishan Chand and Gopal-Krishna]{{ \large Krishan Chand$^{1,2}$\thanks{E-mail: krishan@aries.res.in (KC); gopaltani@gmail.com (G-K)} and 
Gopal-Krishna$^{3}$}\\\\
$^{1}$Aryabhatta Research Institute of Observational Sciences (ARIES), Manora Peak, Nainital 263002, India\\
  $^{2}$Department of Physics, Kumaun University, Nainital 263002, India\\
  $^{3}$UM-DAE Centre for Excellence in Basic Sciences, Vidyanagari, Mumbai-400098, India\\
}
\begin{document}
\date{Accepted ---; Received ---; in original form ---}

\pagerange{\pageref{firstpage}--\pageref{lastpage}} \pubyear{2022}

\maketitle

\label{firstpage}
\begin{abstract}
Flat-spectrum radio quasars (FSRQs) whose brightness is dominated by a relativistically beamed core, are frequently found in `blazar state' commonly inferred from a high optical polarization ( > 3\%), and/or a large continuum variability. Here we use these two prime optical markers to investigate continuance of an FSRQ in  blazar (or non-blazar) state over an exceptionally long time baseline spanning 4 decades. Our basic sample is a well-defined, unbiased set of 80 FSRQs whose blazar state stood confirmed during 1980s from optical polarimetry. Four decades later, blazar state of each FSRQ is ascertained here from variability of their optical light-curves of typical duration $\sim$ 3.5 years, a low noise (rms $\sim$ 2\%) and good cadence ($\sim$ 3 days), obtained under the Zwicky Transient Facility project ongoing since 2018. For about 40\% of these FSRQs, blazar state could be ascertained additionally from the opto-polarimetric survey RoboPol (2013-2017). From both these databases it is found that only $\sim$ 10\% of the FSRQs have undergone a blazar $\leftrightarrow$ non-blazar state transition over the past 3 - 4 decades. This reinforces the case for a long-term stability of blazar state in individual FSRQs, despite their state fluctuating more commonly on year-like time scales.

 \end{abstract}


\section{Introduction}
\label{Introduction}
Early studies of QSOs unveiled a polarimetric distinction between the overwhelming majority ($\sim$ 99\%) of optically-selected QSOs and the remaining $\sim$ 1\% of them which showed a high fractional optical polarization ($p_{opt}$ $>$ 3\%, \citealp{Stockman1984ApJ...279..485S}). This tiny radio-loud minority also resembled BL Lac objects in their optical and radio continua \citep{Stein1976ARA&A..14..173S}. Although apparently arbitrary, this polarization threshold has since become the thumb-rule for separating High-Polarization-Quasars (HPQs, which are nearly always radio-loud) from Low-Polarization-Radio-Quasars (LPRQs). It was also found that a compact (parsec-scale) radio core exhibiting a flat/inverted radio spectrum owing to synchrotron self-absorption, is a distinctive feature of HPQs. The term `blazar' was adopted \citep{Angel1980ARA&A..18..321A,Moore1981ApJ...243...60M} for HPQs, jointly with BL Lacs which too exhibit a strong and variable polarization, but differ from HPQs in having very weak optical/UV emission lines. 
In both, the dominant contribution to the observed highly variable emission (continuum and polarised) at centimetre and shorter wavelengths comes from a jet of nonthermal emission relativistically beamed towards us \citep{Blandford1978PhyS...17..265B}.

From the radio perspective, a flat/inverted radio spectrum of a quasar, which marks the dominance of its radio core, has often been deemed adequate for assigning a blazar designation. For instance, it was asserted in \citet{Impey1990ApJ...354..124I}, hereafter IT90, that essentially every quasar whose radio emission is dominated by its core, exhibits a prominent blazar component at optical wavelengths \citep[see, also, ][]{Wills1992ApJ...398..454W,Maraschi2003ApJ...593..667M,Meyer2011ApJ...740...98M}. The criterion of strong linear polarization ($p_{opt}$ $>$ 3\%) as the marker of prominent optical synchrotron component has been employed to identify blazars among flat-spectrum radio quasars (FSRQs), with a success rate of 40 - 50\% 
(\citealp{Fugmann1988AAS...76..145F}; \citealp{Wills1989LNP...334..109W}; IT90). Actually, this is an underestimate since variability of polarization being a key attribute of blazars, polarization of some genuine blazars may sometimes dip below the 3\% threshold, resulting in their mis-classification as an LPRQ.

The issue of duty cycle of high optical polarization state (i.e., blazar state) among FSRQs was examined, e.g., by \citet{Fungmann1988AA...205...86F} and IT90 and it was estimated that about 2/3rd of FSRQs have $p_{opt}$ $>$  3\% some of the time. In fact, it has been suggested that all quasars with a dominant compact radio core have $p_{opt}$ $>$ 3\%, at least some of the time and duty cycle in this HPQ (i.e., blazar) state increases with radio compactness (\citealp{Fungmann1988AA...205...86F}; \citealp{Impey1991ApJ...375...46I}, hereafter ILT91; also, \citealp{Lister2000ApJ...541...66L}). While bearing in mind the possibility that a blazar's polarization might occasionally dip below $p_{opt}$ = 3\% \citep[e.g.,][]{Moore1984ApJ...279..465M,Lister2000ApJ...541...66L}, the division at $p_{opt}$ = 
3\% (see above) to discriminate between LPRQs and HPQs, as also adopted here, remains largely valid \citep[see, e.g.,][and references therein]{Algaba2011MNRAS.411...85A}. Exceedingly rare cases of QSOs, such as PHL 5200, are known which have $p_{opt}$  > 3\% (due to scattering, see \citealp{Stockman1981ApJ...243..404S}), but lack radio detection \citep[e.g.,][]{Gopal-Krishna1980A&A....90L...1G}.
Note also that some blazars may never appear as HPQs, e.g., nucleus of the intensively monitored quasar 3C 273 is found to harbour a mini-blazar, but its expected variable contribution is almost never able to push $p_{opt}$ beyond the $3\%$ threshold \citep{Courvoisier1988Natur.335..330C,Impey1989ApJ...347...96I,Wills1989LNP...334..109W}. 

Variable flux density is another fundamental characteristics of blazars and their polarised and continuum optical emissions are known to vary on time scales which can be as short as minutes (e.g., reviews by \citealp{Wagner1995ARA&A..33..163W}; \citealp{Ulrich1997ARA&A..35..445U}; \citealp{Marscher2016Galax...4...37M}; \citealp{Gopal2018BSRSL..87..281G}). Variable optical continuum, like the fractional polarization, has a strong statistical link to blazar activity, underscoring that a high polarization is associated with the variable component \citep{Moore1984ApJ...279..465M}. ILT91 have highlighted the near-identical dependence of polarization and optical continuum variability on radio compactness (on VLBI scale) which is a proxy for the relativistically beamed synchrotron component. Their inference was partly based on the nearly 2-decades long monitoring campaigns to measure optical variability of over 100 AGN on year-like time scales \citep{Pica1988AJ.....96.1215P,Webb1988AJ.....95..374W}. However, as noted by ILT91, this time scale is sub-optimal since optical polarization had been observed to vary on much shorter (week-like) time scale. 
This limitation of sensitivity and cadence of optical light curves has been largely overcome in the present study.

Intra-night optical variability (INOV) is yet another, key attribute of blazars, with variability amplitudes ($\psi$) above 3-4\% occurring with a duty cycle of around 40-50\% \citep[][and references therein]{Gopal2018BSRSL..87..281G}. The dependence of INOV on $p_{opt}$ was investigated by \citet{Goyal2012AA...544A..37G} in an extensive campaign of intranight optical monitoring of 9 HPQs and 12 flat-spectrum LPRQs. Whereas strong 
INOV ($\psi$ $>$ 4\%) was detected for the HPQs on 11 out of 29 nights, it was seen for the LPRQs on just 1 out of 44 nights. This striking contrast demonstrated that INOV is more fundamentally linked to optical polarization than it is to relativistic beaming of the nuclear radio jet. Rather unexpectedly, the tight correlation of INOV with $p_{opt}$ was observed even though the polarimetric classification of the FSRQs sub-populations monitored by them had been done more than two decades prior to their INOV campaign. Although this strong correlation only reflects long-term ensemble behaviour of the two FSRQ sub-populations monitored by them, it does not seem to tie up with the findings of IT90, based on optical polarimetric measurements of 41 quasars at two epochs with a median separation of $\sim$ 1 year, in which almost a quarter  of the quasars were found to switch their polarization state (LPRQ to HPQ, or vice versa). Here we endeavour to probe the issue of persistence of polarization state of individual FSRQs, employing a larger quasar sample, together with optical continuum variability (on week to year-like time scales) as an alternative, well-established tracer of blazar state \citep[e.g.,][]{Bauer2009ApJ...699.1732B}. 
For this, we make use of the optical light-curves (LCs) of high sensitivity and cadence, obtained for a large number of quasars, under the ongoing ZTF project launched in 2018 \citep{Belim2019PASP..131a8002B}.

\floatsetup[table]{capposition=top}
  \begin{table*}
  \centering
\begin{minipage}{163mm}
    \caption{The basic radio/optical data for sample 1, together with information derived from the ZTF optical light-curves \citep{Belim2019PASP..131a8002B}  and the opto-polarimetric survey RoboPol \citep{Blinov2021MNRAS.501.3715B}. The full table 1a (sample 1) and table 1b (sample 2) are available in the on-line material.}
    \label{tab1}
    \resizebox{\textwidth}{!}{%
\begin{tabular}{ccccccccccccccc}
  \hline\\
\multicolumn{1}{c}{Source} &\multicolumn{1}{c}{Other name} &\multicolumn{1}{c}{{\it z}}  &\multicolumn{1}{c}{Flux at} & Radio   &\multicolumn{1}{c}{App. mag.} &\multicolumn{1}{c}{App. mag.}&\multicolumn{1}{c}{$p_{opt}$} &Pol.& \multicolumn{1}{c}{N (LC)} &Total duration &$\sigma_{med}$  &      &Variability        &${p_{opt}}$ (\%)\\
 SDSS name                &                                &                            &{5 GHz}                     &Sp.Index &(SIMBAD)                      &(SDSS)                       &(\%)                          &class&                            &(T) of the LC  &                &      &type & RoboPol\\
                           &                               &                            &      (Jy)                  &$\alpha$ &{\textcolor {blue}B}          &                             &                              && 		                &$T_{obs}$, ($T_{int}$)      &                 &	&                   &${p1}$ (JD*)  \\
                           &                               &                            &                           &($f_{\nu} \propto \nu^{\alpha}$) &{\textcolor {cyan}V}&{\textcolor {Green}g}&                         &&{\textcolor {Green}g}       & (days)        &                 &${\textcolor {Green}{\Delta m_g}}$&           &${p2}$ (JD*)    \\
                           &                               &                             &                         &           &{\textcolor {red}R}   &{\textcolor {red}r} &                                               &&{\textcolor {red}r}         &               &                  &${\textcolor {red}{\Delta m_r}}$	&           &${p3}$ (JD*)    \\
&                               &                             &                         &           &   & &                                               &&         &               &                  &	&           &mean ${p}$, (N)   \\
(1)                        &   (2)                   &(3)                   &(4)                      & (5)     &(6)                     &(7)                    &(8)      & (9)                     &(10)                        &(11)      &(12)               &(13)       & (14) & (15)\\\\
 \hline\\
0106+013             &4C 01.02     & 2.107        &2.41   &0.12&{\textcolor {blue}{18.54}}     &           &${2.2\pm1.1}$ { (7.1)$^{a}$}  &HPQ   &       &  &   &  &V (Mt)     &\\
J010838.77+013500.32 &          & 2.099  &       &&{\textcolor {cyan}{18.39}}     & {\textcolor {Green}{18.29}}    &                 & & {\textcolor {Green}{235}}  & {\textcolor {Green}{1251.76 (403.92)}}&{\textcolor {Green}{0.036}}  &{\textcolor {Green}{1.88}}         &       &\\
                     &             &         &       && {\textcolor {red}{19.05}}    & {\textcolor {red}{18.10 }}               &      &  & {\textcolor {red}{238}} & {\textcolor {red}{1261.68 (407.12)}}&{\textcolor {red}{0.034}}   &{\textcolor {red}{2.35}}         & &\\\\
0133+476             &  OC 457           &0.860         &2.92  &0.57 &{\textcolor {blue}{---}}    &	            & ${20.8\pm0.7}$ { (20.8)$^{a}$}    &HPQ   &	     & &  &  &  V (Mt)   &$14.6\pm0.8$ (6591)\\     
J013658.59+475129.10 &             & 0.859  &       & &{\textcolor {cyan}{18.00}}    &{\textcolor {Green}{18.70}}           &  ${20.8\pm0.7}$ { (20.8)$^{b}$}     &               &{\textcolor {Green}{271}}    & {\textcolor {Green}{1257.77 (676.58)}} &{\textcolor {Green}{0.100}}    &{\textcolor {Green}{2.46}}         &      &$6.4\pm1.5$ (7617)\\
&             &         &       & &{\textcolor {red}{19.25}}    &{\textcolor {red}{18.13}}           &             &       & {\textcolor {red}{423}} &{\textcolor {red}{1260.88 (678.26)}}&{\textcolor {red}{0.067}}  &     {\textcolor {red}{2.79}}        &      &$4.1\pm1.5$ (7607)\\
              &             &         &       & &    &          &             &       &  &&  &             &      &$<8.75>$, (N=64)\\

---            &---       & ---        &---   &---&---    &---           &---&---        &---	&---	&---   &---  & ---     &\\
\hline\\
\multicolumn{15}{l}{{\bf Col. (3):} Redshift, the upper value is from \citet{Impey1990ApJ...354..124I} and the lower value is from NED; {\bf Col. (4):} Flux density at 5 GHz, taken from \citet{Impey1990ApJ...354..124I}; {\bf Col. (5):} Radio spectral index is taken from} \\
\multicolumn{15}{l}{ \citet{Kuehr1981AAS...45..367K}; {\bf Col. (8):} The `first pass' measured value of optical polarization given by (a) \citet{Impey1990ApJ...354..124I} and (b) \citet{Impey1991ApJ...375...46I}; their maximum measured value ($p_{opt}$(max)) is given in parentheses;}\\
\multicolumn{15}{l}{ {\bf Col. (9):} polarization class of the source ($p_{opt}$(max) $<$ 3\% for LPRQs and $>$ 3\% for HPQs). The polarization class in parentheses is for those three sources for which the quoted one-sigma error on $p_{opt}$(max) would}\\
\multicolumn{15}{l}{ push the source to the other polarization class; {\bf Col. (10):} Number of points in the ZTF light-curve; {\bf Col. (11):} The value in parentheses is the intrinsic duration (days) of the ZTF light-curve, i.e., total duration}\\
\multicolumn{15}{l}{(T / (1+$z_{NED}$);  {\bf Col. (12):} Median value of the rms errors of the N points in the ZTF light-curve; {\bf Col. (13):} Total (peak-to-peak) variation of apparent magnitude in the ZTF light-curve (eq. \ref{eq:1}); {\bf Col. (14):} The assigned}\\
\multicolumn{15}{l}{variability type (see section \ref{The quasar samples}); {\bf Col. (15):} RoboPol measured $p_{opt}$, taken from \citet{Blinov2021MNRAS.501.3715B} (see section \ref{Discussion}). The JD in parentheses marked with `*' corresponds to Julian date 
minus 2450000 and the } \\
\multicolumn{15}{l}{ parameter N in parentheses corresponds to the number of polarization measurements. The terms {$p1$}, {$p2$} and {$p3$} are explained in section \ref{Discussion}}.\\

\end{tabular}
 } 
 \end{minipage}
\end{table*}

\section{The quasar samples and optical variability }
\label{The quasar samples}
The two samples of radio-loud quasars studied here are derived from the opto-polarimetric study of IT90, which is primarily based on two statistically complete sets of quasars drawn from the 1-Jy catalogue of 518 sources, defined at 5 GHz by \citet{Kuehr1981AAS...45..367K}. Their first set consists of 90 quasars stronger than 2 Jy at 5 GHz and the second set contains 50 FSRQs ($\alpha$ (2.7 - 5 GHz) $>$ -0.5) having flux densities between 1.5 and 2 Jy at 5 GHz. For both sets we downloaded on 24-March-2022 the LCs of the ZTF survey\footnote{The database of the ongoing ZTF project, \url{https://www.ztf.caltech.edu}}, in the more sensitive r and g bands, after excluding the two quasars for which $p_{opt}$ was not given in IT90, as well as quasars for which either (i) a ZTF LC was not available in even one of the two bands, or (ii) the available LC in neither band has even 25 measurements (N $<$ 25), or (iii) the LCs are very noisy (i.e., median $\sigma$ for the data points is $>$ 0.15-mag). These exclusions from the first set of IT90 led to our sample 1 (60 quasars, including 48 FSRQs) and from their second set led to our sample 2 (30 quasars, all FSRQs). Basic optical/radio data for each quasar in these two samples are provided in Tables \ref{tab1}a \&\ref{tab1}b, together with relevant information on their ZTF LCs and the optical variability parameters derived from them, as well as the published optical polarimetric data taken from IT90 and the recent opto-polarimetric survey RoboPol \citep{Blinov2021MNRAS.501.3715B}. 
For the LCs in both our samples, the on-line Fig. S1 displays the histograms of time duration (observed and rest-frame) and $\sigma$ (median) (typically, $\sim$ 0.02-mag and $\sim$ 0.03-mag for the LCs in r-band and g-band, respectively). The rest-frame (i.e., intrinsic) time duration, $T_{int}$  is $>$ 1 year for 90 - 95\% of the LCs. Excellent consistency is found between the LCs in r and g bands, but since the two often do not fully overlap in time, the temporal coverage for many quasars has been augmented by considering the two LCs jointly. For deriving the peak-to-peak variability amplitude ($\Delta m$) of a quasar, we  have used the LC of the band which shows a higher variability amplitude ($\Delta m$), where $\Delta m$ is given by \citep{Heidt1996A&A...305...42H,Romero1999A&AS..135..477R}:
\begin{equation} \label{eq:1}
 \Delta m= \sqrt{({A_{max}}-{A_{min}})^2-2\sigma^2}\\   
\end{equation}
Here $A_{max}$ and $A_{min}$ are the maximum and minimum values in the LC and $\sigma^2$ is the mean square rms error for the data points.

\begin{figure}

  \includegraphics[width=9.51cm,height=12.00cm,trim=0.10cm 0cm 0.0cm 1.7cm,clip]{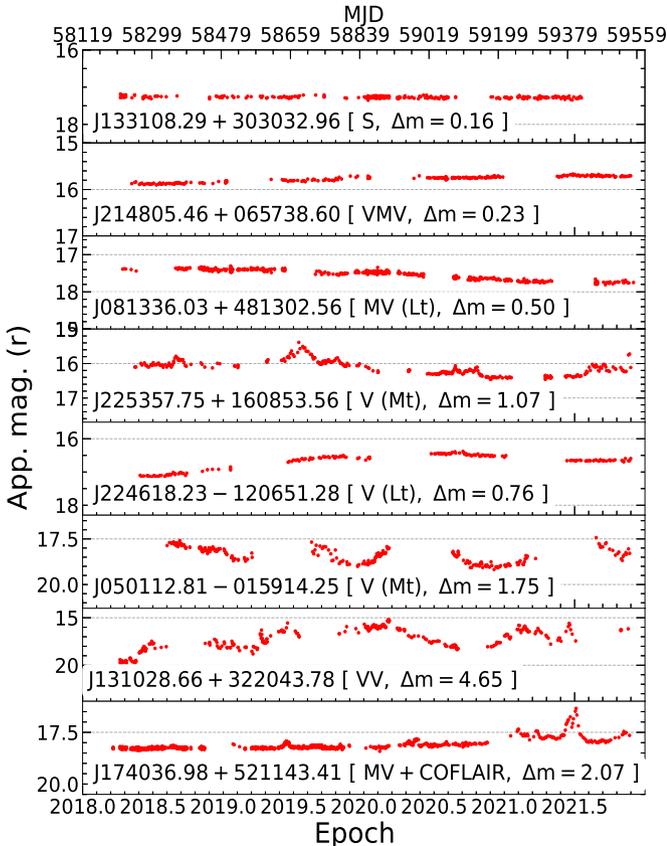}\\
  \vspace{-0.42in}
  \caption{Representative light-curves (r-band) for the different variability types (Section \ref{The quasar samples}).}
\label{fig:2} 
\end{figure}

For each quasars in sample 1 and sample 2, we have assigned a variability type, based on our visual inspection of its LCs (column 14 in Table  \ref{tab1}). Although, we do not use this (subjective) classification for statistical purpose, it is retained here as a useful global descriptor of continuum variability,  both its amplitude and time-scale (on-line Figs. 1a-60a \& 1b-30b). The adopted classifications for variability time-scale are:  Medium-term (Mt) for week/month-like and Long-term (Lt) for year-like. The 5 variability types assigned to the LCs are: (1) `Steady' or `Very Mildly Variable' (S / VMV); (2) `Mildly Variable' (MV); (3) `Variable' (V); (4) `Violently Variable' (VV) and (5) aperiodic `CONfined FLARE(s)' (CONFLARE), superposed on a relatively quiescent LC of S/VMV, or MV type. In these LCs, the S/VMV type variability is always found to be long-term (Lt) only, while the V \& VV type LCs always exhibit medium-term (Mt) variability. Nearly always, these variability types are found to have: $\Delta m$ $<$ 0.5-mag (S/VMV), 0.5 - 1.0-mag (MV), 1.0 - 3.0-mag (V) and  $> $3-mag (VV).  Representative LCs for each variability type in our samples are shown in Figure \ref{fig:2}.


\section{Results}
\label{Results}
Tables  \ref{tab1}a \&\ref{tab1}b (column 8) reproduce from IT90 the `first pass' value and the maximum known value of $p_{opt}$ for each quasar, out of the available measurements (at most a few, for most of the quasars). Often, the two values listed for a quasar in IT90 are identical, since just one polarimetric measurement was available then. In view of the small number of measurements available for individual quasars, we have preferred for polarimetric classification the use of $p_{opt}$ (max), instead of mean polarization, in order to reduce the possibility of mis-identifying genuine HPQs as LPRQs, owing to polarization variability of HPQs. Thus, for the quasars in our two samples, with HPQ designation, a blazar-state had been identified in the 1980s, based on optical polarimetry (Tables \ref{tab1}a \&\ref{tab1}b). Now, i.e., after an interval of 4 decades, we can check for each of the quasars {\it individually} the status of its blazar activity, on the basis of the amplitude of optical continuum variability in its ZTF light-curves (on-line Figs. 1a-60a \& 1b-30b). Conversely, we can also look for the quasars whose low polarization ($p_{opt}$ (max) $<$ 3\%) during the 1980s indicated a non-blazar state then, but which have now entered a blazar state, as signified again by a strong variability in their ZTF light-curves. For this check, we extracted from Tables \ref{tab1}a \& \ref{tab1}b, two categories of quasars which are candidates for transition to the opposite state, i.e., either from a confirmed blazar state in 1970-80s to a clearly quiescent (i.e., non-blazar) state now, or, vice versa, during the intervening $\sim$ 4 decades. To facilitate a quantitative estimation, we adopt the following reasonably robust criteria to define these two types of transitions as: (i) blazar to non-blazar state ($p_{opt} (max)$ $>$ 3\% to  $\Delta m$ $<\sim$ 0.5-mag), and (ii) non-blazar to blazar state ($p_{opt}$ (max) $<$ 3\% to $\Delta m$ $>\sim$ 1.0-mag). Comments on the quasars short-listed under these two categories of state transition are given below (see, also, Tables \ref{tab1}a \& \ref{tab1}b):\\

{\bf Category I:} Blazar to non-blazar state: The 3 candidates are: \\

J0405-1308: The two available measurements prior to 1990 gave $p_{opt} = 20.2\%$ and  $3.8\pm0.5\%$, demonstrating its blazar state. The observed mild optical variability ($\Delta m = 0.45\pm0.02$) confirms  this FSRQ as a strong case of state transition.

J0823+2223: With $p_{opt} = 5.2 \pm 1.2\%$ (IT90), the source was a known HPQ in 1980s. However, its ZTF LCs are short ($T_{int}$ $\sim$ 5 months) and poorly sampled, with a large gap. The possibility that its observed $\Delta m$ of 0.37$\pm0.02$ may hence be seriously underestimated, renders this blazar a weak candidate for state transition.

J0957+5522: With $p_{opt} = 9.6\pm1.8\%$ (IT90) and $6.4\pm0.5\%$ (ILT91), this FSRQ was a bona-fide blazar in 1980s. Its well-sampled ZTF light-curves give a low $\Delta m = 0.53\pm0.03$ (g-band) and $0.45\pm0.02$ (r-band), making it a strong case of state transition.\\

{\bf Category II:} Non-blazar to blazar state:\\

J0530+1331: Its designation as LPRQ (i.e., non-blazar) in 1980s is based on the single available measurement ($p_{opt} = 0.3\pm1.0\%$, \citealp{Fugmann1988AAS...76..145F}). So, this could even be a case of an HPQ masquerading as LPRQ. Secondly, although its $\Delta m$ of 1.43-mag seems large, its LCs in both r and g bands are quite noisy and also poorly sampled (on-line Table 1a; on-line Fig. 13a), rendering it a weak case for state transition.

J0750+1231: It was a bona-fide non-blazar (LRPQ) based on the two measurements available in 1980s, giving $p_{opt} = 1.0\pm1.0\%$ (IT90) and $2.1\pm1.1\%$ \citep{Wills1992ApJ...398..454W}. The observed $\Delta m$ of 1.43-mag is based on good quality ZTF light-curve (on-line Fig. 18a), making it a robust case of transition to blazar state, which is further reinforced by the relatively recent RoboPol measurements yielding ${p_{max} = 12.3\pm1.5\%}$ (on-line Table 1a).

J0956+2515: The available two early measurements, giving $p_{opt} = 0.7\pm0.4\%$ (IT90) and $2.2\pm0.8\%$ \citep{Moore1984ApJ...279..465M}, confirmed its non-blazar state in 1980s. The high $\Delta m$ of 1.60-mag, based on good-quality ZTF LCs (on-line Fig. 19b; on-line Table 1b), makes it a strong case of blazar state transition.

J1357-1527: Its identification in 1980s as a non-blazar (LPRQ) rests on single measurement ($p_{opt} = 1.4\pm 0.5\%$, IT90), which therefore does not rule out its being an HPQ in 1980s. Thus, inspite of a strong variability ($\Delta m$ = 1.14-mag, on-line Table 1b), based on good quality ZTF LCs (on-line Fig. 22b), this FSRQ remains an uncertain case of state transition.

J1557-0001 : Given that its LPRQ (non-bazar) classification in the 1980s rests on single measurement ($p_{opt} = 1.2\pm1.3\%$, IT90), there is a non-negligible chance of its being a HPQ then. Thus, despite a strong variability ($\Delta m$ = 1.66-mag, on-line Table 1a) seen in its high-quality ZTF LC (on-line Fig. 41a), this FSRQ remains a weak case of state transition.

J1635+3808 : The old designation of this FSRQ as a non-blazar (LPRQ) is secure, based on 3 available measurements during 1980s: $p_{opt} = 2.6\pm1.0\%$ \citep{Moore1984ApJ...279..465M}, $1.1\pm0.2\%$ (ILT91) and $0.8\pm0.9\%$ \citep{Wills1992ApJ...398..454W}. Thus, the high variability ($\Delta m$ = 2.67-mag) seen in its good-quality ZTF  LCs establishes it as a strong case of state transition. This transition was already highlighted in \citet{Lister2000ApJ...541...66L} and confirmed in its 
RoboPol polarimetry, giving $p_{opt}$ (median) =  $9.9\pm0.1\%$ \citep{Angelakis2016MNRAS.463.3365A}.

\section{Discussion}
\label{Discussion}
From the above analysis, it is seen that out of the 49 quasars in Tables \ref{tab1}a \& \ref{tab1}b, which had been established (through optical polarimetry) as HPQs during mid-1970 to mid-1980s, only two (i.e., 4\%) show robust signs of transition to non-blazar state in their ZTF LCs obtained around 2020. The corresponding numbers for the reverse transition, from LPRQ (i.e., non-blazar) to blazar state, are 3 quasars out of 41 (7.3\%). Even if, the 10 steep-spectrum ($\alpha$ $<$ - 0.5) LPRQs are excluded, the transition fraction for the remaining 31 LPRQs would still be only 9.7\%. Thus, from this source-wise analysis it is evident that despite the elapsing of close to 4 decades, ${< \sim 10\%}$ of the FSRQs show clear signs of change in their blazar state in either direction. A broadly similar inference, albeit on the basis of ensemble behaviour, was reached in \citet{Goyal2012AA...544A..37G}, in their intranight optical variability (INOV) campaign around 2006
(section \ref{Introduction}) which showed that the memory of polarization state identified in 1980s had endured and got strongly reflected in the INOV behaviour about 2 decades later. Here we find a similarly high level of persistence of blazar (or, non-blazar) state over twice longer time baseline (i.e., 4 decades), by checking the blazar state of each FSRQ {\it individually}.  
How then to reconcile these results with the finding of IT90 that a significant fraction of FSRQs changes its polarization state on year-like time scales (section \ref{Introduction})? 

To further scrutinise this point, we shall now subject our quasar sample (Tables \ref{tab1}a \& \ref{tab1}b) to an additional check in which the blazar status of a quasar at {\it both} ends of the time baseline is determined using optical polarimetric data alone (this option entails a substantial reduction in the sample size, however). For this, we focus attention on those FSRQs in our samples 1 \& 2 extracted from the polarimetric survey of IT90, which are also covered in the RoboPol survey conducted during 2013-2017, i.e., after an interval of nearly 3-4 decades  \citep{Blinov2021MNRAS.501.3715B}.
Out of the 90 quasars in our two samples, 10 are LPRQs of steep radio spectrum (the 2 HPQs, although having moderately steep radio spectra, will be deemed as FSRQs and clubbed with them). The remaining 80 quasars (FSRQs) include 49 HPQs and 31 LPRQs. Thirty-two of these FSRQs have been observed in the RoboPol survey, mostly multiple times (median N = 21). A proper source-wise comparison of these measurements with the old polarimetric data provided in IT90, should be based on similar number of measurements in the two data sets.
Since IT90 reports no more than a few polarimetric measurements for the vast majority of their sources, the comparison sample has been derived here from the RoboPol database by limiting to 3 
measurements per source (N = 3). For sources with > 3 RoboPol measurements, we selected 3 values using a random number generator. The values based on this unbiased selection process are listed in column 15 of Table \ref{tab1}, together with the total number of available measurements (N) and their average value of $p$ as published in the RoboPol catalogue. Comparison of these opto-polarimetric measurements with the old opto-polarimetric measurements (columns 8 \& 9) reveals the following strong cases of polarization state transition:
(i) J0405-1308: HPQ to LPRQ transition (consistent with the ``mildly variable'' classification of its ZTF light-curve); (ii) J1635+3808: LPRQ to HPQ transition (also mentioned in section \ref{Results}), again consistent with the ``variable'' classification of its ZTF LC). Besides these two confirmed cases, J0750+1231 may have also undergone transition (from LPRQ to HPQ state), however its LPRQ classification in IT90 is based on just one measurement, which leaves a significant chance that it was a case of HPQ masquerading as LPRQ. Thus, in summary, a comparison of the IT90 and RoboPol opto-polarimetric surveys independently confirms that at most 2-3 (i.e., < 10\%) out of the 32 FSRQs common to both surveys have undergone a change of polarimetric state during the intervening 3 - 4 decades. This independent estimate of state change, based purely on polarimetric data, is fully in accord with our above estimate using a 2.5 times larger sample of FSRQs and comparing their blazar/non-blazar states as inferred from the polarimetric (IT90) and photometric (ZTF) surveys separated by 4 decades. It is thus, remarkable that >$\sim$ 90\% of the FSRQs which had been identified as HPQs in the 1970-80s (IT90) have retained their blazar mode for nearly 4 decades, as inferred using the ZTF light-curves (onwards 2018) and corroborated using the polarimetric survey RoboPol during 2013-2017. The fluctuating polarization state of around a quarter of FSRQs on year-like time scales, as reported in IT90 and also seen using our afore-mentioned 32 FSRQs covered in the RoboPol survey 
(see below), does not seem to swamp the long-term stability of opto-polarimetric state which is found to persist for at least a few decades. Conceivably, the observed short-term polarization fluctuations may be related to transient events, like the formation/ejection of radio knots mapped on VLBI scales and probably manifested by $\gamma$-ray flares \citep[e.g.,][]{Marscher2008Natur.452..966M,Alok2017MNRAS.472..788G}, which also seem to have a year-like occurrence rate, on an average \citep{Savolainen2002A&A...394..851S,Lister2009AJ....138.1874L,Liodakis2018MNRAS.480.5517L}. 

Finally, in order to quantify the occurrence of polarization state transition on year-like time scales, we have selected out of the present sample of 32 RoboPol FSRQs all 27 FSRQs for which N > 1. The on-line Fig. S2 shows for these 27 FSRQs a plot of $p_{max}$ against $p_{min}$, for which the time interval is found to have a median value of 1.1 yr (see column 15 of Table \ref{tab1}). It is evident that 7 of the 27 FSRQs ($\sim 26\%$) crossed the $p_{opt}$= 3\% threshold on year-like time scales, confirming an early independent estimate by IT90. Note, however, that the actual fraction may be somewhat higher since another 7 FSRQs are border-line cases (i.e., within 1$\sigma$ error of  $p_{opt}$= 3\%), some of which could be genuine transitions.

\section{Conclusions}
Since the 1980s it has been argued  that at a given time, about 2/3rd of a sample of FSRQs is in a blazar state which is marked by a high level of optical polarization ( > 3\%) and flux variability. Here we have endeavoured to estimate for how long an individual FSRQ, upon entering the blazar state, remains in that state. Although, optical polarimetry had hinted that nearly a quarter of FSRQs undergo a change of state on year-like time scales, intranight optical variability observations indicated that memory of blazar state persists for at least an order-of-magnitude longer time span (section \ref{Introduction}). We have examined this apparent dissonance by studying a well-defined sample of 80 FSRQs whose blazar state was determined during 1980s through optical polarimetry (and confirmed for 49 of them) and the same has now been checked by us by measuring their optical continuum variability in the high quality light-curves obtained in the ongoing Zwicky Transient Factory project launched in 2018. We find that only $\mathbf{<\sim}$ 10\% of the FSRQs have changed their blazar state over the 4-decades long time span. Furthermore, this estimate is found to be reinforced by the opto-polarimetric survey RoboPol (2013-2017) which covers ${\sim}$ 40\% of our FSRQ sample. This leads us to conclude that typically, blazar state of an FSRQ is likely to persist for at least a few decades, although state transitions on year-like time scales may also occur in some cases, probably associated with short-term processes, like the formation and ejections of relativistic plasma blobs (VLBI knots) from the active nucleus.
\section*{Acknowledgments}
GK acknowledges a Senior Scientist fellowship from the Indian National Science Academy.
This work is based on observations obtained with the Samuel Oschin Telescope 48-inch and the 60-inch Telescope at the Palomar Observatory as part of the Zwicky Transient Facility project. ZTF is supported by the National Science Foundation under Grants No. AST-1440341 and AST-2034437 and a collaboration including current partners Caltech, IPAC, the Weizmann Institute for Science, the Oskar Klein Center at Stockholm University, the University of Maryland, Deutsches Elektronen-Synchrotron and Humboldt University, the TANGO Consortium of Taiwan, the University of Wisconsin at Milwaukee, Trinity College Dublin, Lawrence Livermore National Laboratories, IN2P3, University of Warwick, Ruhr University Bochum, Northwestern University and former partners the University of Washington, Los Alamos National Laboratories, and Lawrence Berkeley National Laboratories. Operations are conducted by COO, IPAC, and UW. 
This research has made use of data from the RoboPol programme, a collaboration between Caltech, the University of Crete, IA-FORTH, IUCAA, the MPIfR, and the Nicolaus Copernicus University, which was conducted at Skinakas Observatory in Crete, Greece.
This work has also made use of the NASA/IPAC Extragalactic Database (NED) which is operated by the Jet Propulsion Laboratory, California Institute of Technology, under contract with the National Aeronautics and Space Administration.

\section*{Data availability}
The data used in this study are publicly available in ZTF DR8.


\bibliography{references}
\label{lastpage}
\end{document}